\documentclass[conference]{IEEEtran}
\IEEEoverridecommandlockouts
\usepackage{amsmath,amssymb,amsfonts}
\usepackage{algorithmic}
\usepackage{graphicx}
\usepackage{xspace}
\usepackage{textcomp}
\usepackage[inline]{enumitem}
\def\BibTeX{{\rm B\kern-.05em{\sc i\kern-.025em b}\kern-.08em
    T\kern-.1667em\lower.7ex\hbox{E}\kern-.125emX}}

\makeatletter
\newcommand*\titleheader[1]{\gdef\@titleheader{#1}}
\AtBeginDocument{%
  \let\st@red@title\@title
  \def\@title{%
    \bgroup\normalfont\small\centering\@titleheader\par\egroup
    \vskip1.5em\st@red@title}
}
\makeatother

\newcommand{\exanestgrant}{Grant Agreement No. 671553 (ExaNeSt)\xspace}

\newcommand{\hbp}{HBP\xspace}
\newcommand{\hbpsgafirst}{Grant Agreement No. 720270 (HBP SGA1)\xspace}

\newcommand{\ie}{\textit{i.e.}\xspace}
\newcommand{\eg}{\textit{e.g.}\xspace}

\newcommand{\dpsnn}{DPSNN\xspace}

\newcommand{\dpsnnlong}{Distributed and Plastic Spiking Neural Network\xspace}	

\newcommand{\galileo}{GALILEO\xspace}
\newcommand{\cineca}{CINECA\xspace}

\title{Gaussian and exponential lateral connectivity on distributed spiking neural network simulation\\
}
\titleheader{Published in final peer reviewed version as: E. Pastorelli et al., "Gaussian and Exponential Lateral Connectivity on Distributed Spiking Neural Network Simulation," 2018 26th Euromicro International Conference on Parallel, Distributed and Network-based Processing (PDP), Cambridge, 2018, pp. 658-665. doi: 10.1109/PDP2018.2018.00110}

\author{
\IEEEauthorblockN{Elena Pastorelli}
\IEEEauthorblockA{INFN Sezione di Roma and PhD Program in Behavioural Neuroscience, ``Sapienza'' University of Rome\\
  Rome, Italy\\
  elena.pastorelli@roma1.infn.it}
\\
\IEEEauthorblockN{Pier Stanislao  Paolucci, Francesco Simula, Andrea Biagioni, Fabrizio Capuani, Paolo Cretaro, Giulia De Bonis, \\Francesca Lo Cicero, Alessandro Lonardo, Michele Martinelli, Luca Pontisso, Piero Vicini}
\IEEEauthorblockA{INFN Sezione di Roma \\
  Rome, Italy\\
\{name.surname\}@roma1.infn.it}
\\
\IEEEauthorblockN{Roberto Ammendola}
\IEEEauthorblockA{INFN Sezione di Tor Vergata and Electronic Engineering Dept., University of Roma ``Tor Vergata''\\Rome, Italy\\
roberto.ammendola@roma2.infn.it}
}

\begin{document}
\maketitle

\begin{abstract}
We measured the impact of \mbox{long-range}
exponentially decaying \mbox{intra-areal} lateral connectivity on the scaling
and memory occupation of a distributed spiking neural network
simulator compared to that of \mbox{short-range} Gaussian
decays.
While previous studies adopted \mbox{short-range}
connectivity, recent experimental neurosciences studies are pointing
out the role of \mbox{longer-range} \mbox{intra-areal} connectivity
with implications on neural simulation platforms.
\mbox{Two-dimensional} grids of cortical columns composed by
up to 11 M \mbox{point-like} spiking neurons with spike frequency adaption were
connected by up to 30 G synapses using short- and
\mbox{long-range} connectivity models.
The MPI processes composing the distributed simulator were run on up
to 1024 hardware cores, hosted on a 64 nodes server platform.
The hardware platform was a cluster of IBM NX360 M5 \mbox{16-core}
compute nodes, each one containing two Intel Xeon Haswell
\mbox{8-core} \mbox{E5-2630} v3 processors, with a clock of 2.40~G Hz,
interconnected through an InfiniBand network, equipped with 4$\times$ QDR switches.
\end{abstract}


\section{Introduction}
\label{sec:dpsnnintro}
We present the impact of the range of intra-areal
lateral connectivity on the scaling of distributed \mbox{point-like}
spiking neural network simulations when run on up to 1024 software
processes (and hardware cores) for cortical models including
tens of billions of synapses.
A simulation including a few tens of billions
of synapses is what is required to simulate the activity of one $cm^2$
of cortex at biological resolution (\eg 54K neuron/$mm^2$ and about
5K synapses per neuron in the rat neocortex area~\cite{schnepel:2015}).
The capability to scale a problem up to such a size allows simulating
an entire cortical area.
Our study focuses on the computational cost of the implementation of
connectivities, pointed out in recent studies reporting about long range 
intra-areal lateral connectivity in many different cerebral areas, 
from cat primary visual cortex~\cite{stepanyants:2009}, to rat 
neocortex~\cite{schnepel:2015, boucsein:2011}, just as examples.
For instance, in rat neocortex, the impact of lateral connectivity on
the pyramidal cells in layer 2/3 and layer 6a, results in $\sim$75\%
of incoming remote synapses to neurons of these layers.

Longer-range \mbox{intra-areal} connectivity
can be modeled by a distance-dependent exponential decay of the
probability of synaptic
connection between pairs of neurons: \ie $ A \cdot exp(\frac{-r}{\lambda}$),
where r stands for the distance between neurons, $\lambda$ is the exponential
decay constant and A is a normalization factor that fixes the total
number of lateral connections.
Decay constants in the range of several hundred microns are required to
match experimental results.

Previous studies considered \mbox{intra-areal} synaptic connections 
dominated by local connectivity: e.g. \cite{schuz:2006} estimated 
at least 55\% the fraction of local synapses, reaching also a ratio of
75\%.
Such \mbox{shorter-range} lateral connectivity has often been modeled
with a distance dependent Gaussian decay~\cite{potjans:2014}
$B \cdot exp(\frac{-r^2}{2\sigma^2}$), where r stands again for
distance between neurons, $\sigma^2$ is the variance that determines
the lateral range and B fixes the total number of projections.
Here we present measures about the scaling of simulations of cortical
area patches of different sizes represented by \mbox{two-dimensional}
grids of ``cortical modules''.
Each cortical module is composed of 1240 \mbox{single-compartment},
\mbox{point-like} neurons (no dendritic tree is represented)
each one receiving up to $\sim$2390 recurrent synapses (instantaneous 
membrane potential charging) plus those bringing external stimuli.
The larger simulated cerebral cortex tile includes 11.4~M
neurons and 29.6~G total synapses.
Exponentially decaying lateral connectivity (\mbox{longer-range}) are
compared to a Gaussian connectivity decay (\mbox{shorter-range}),
analyzing the scaling and the memory usage of
our \dpsnnlong simulation engine (\dpsnn in the following).

On \dpsnn, the selection of the connectomic model has consequences due to:
1) the mapping strategy (neurons and incoming synapses
are placed on MPI processes according to spatial contiguity) and,
2) synaptic messages exchanged between neurons
simulated on different MPI processes entail communication tasks 
among those processes; the higher the number of lateral synaptic connections 
and the longer the interaction distance is, the more intensive the 
communication task among processes becomes.

The impact of other biologically plausible, or experimentally demonstrated, 
connectivity patterns is worth of investigation, but is not covered by this work. 
One of the directions could be the study of the effect of connectivity patterns 
with local modular/clustered connection and global (inter-areal) non-homogenoeus 
lateral connectivity. Such a connectivity has been studied theoretically 
for network dynamical behaviors on a small local neural network~\cite{litwin-KumarDoiron:2012}. 
Experimentally, complex connectivity has been seen mostly for across-area studies,
however see also the emerging strong evidence of local
motifs~\cite{galLondonMarkramSegev:2017}.

The article is structured as follows: Section~\ref{sec:dpsnndesc}
describes the main features of the simulation engine and its
distributed implementation; network models are
summarized in Section~\ref{sec:dpsnnnetwork}, with a 
specific description of the different schemes adopted for the lateral
\mbox{intra-area} connectivity; Section~\ref{sec:dpsnnresult} 
reports the impact of lateral connectivity on the scaling.
A discussion section closes the paper.

\section{Description of the Spiking Neural Network simulator}
\label{sec:dpsnndesc}
The main focus of several neural network simulation projects is the
search for a) biological correctness; b) flexibility in biological
modeling; c) scalability using commodity technology --- \eg
NEST~\cite{gewaltig:2007,kunkel:2017:short},
NEURON~\cite{brette:2007:short}, GENESIS~\cite{NIPS1988_182}.
A second research line focuses more explicitly on computational
challenges when running on commodity systems, with varying degrees of
association to specific platform 
ecosystems~\cite{mattia2000,NAGESWARAN2009791,Izhikevich04032008,Modha:2011}.
Another research pathway is the development of specialized
hardware, with varying degrees of flexibility allowed --- \ie
SpiNNaker~\cite{furber2012}, BrainScaleS~\cite{schmitt:2017:short}.

Instead, the \dpsnn simulation engine is meant to address two 
objectives: (i) quantitative assessment of requirements and 
benchmarking during the development of embedded
\cite{EURETILE:JSA:2016:short} and HPC systems \cite{EuroExa:site}
--- focusing either on network \cite{DSD:EXANEST:2016:short}
or on power efficiency \cite{PAOLUCCI:2015:DPSNN}
--- and (ii) the acceleration of the simulation of specific models
in computational neuroscience --- \eg to study slow waves in large
scale cortical fields~\cite{ruiz:2011,stroh:2013} in the framework
of the \hbp~\cite{hbp:2017:Online} project.

The simulation engine follows a mixed \mbox{time-} and \mbox{event-driven} approach
and implements synaptic \mbox{spike-timing} dependent plasticity 
(~\cite{Gutig:2003,Song:2000}).
It has been designed from the ground up to be natively distributed and
parallel, and should not pose obstacles against distribution and
parallelization on several competing platforms.
Coded as a network of C++ processes, it is designed to be easily
interfaced to both MPI and other (custom) \mbox{Software/Hardware}
Communication Interfaces.

In this work, the neural network is described as a two-dimensional grid of 
cortical modules made up of single-compartment, point-like neurons 
spatially interconnected by a set of incoming synapses.
Cortical modules are composed of several populations of excitatory and
inhibitory neurons.
Cortical layers can be modeled by a subset of those populations.
Each synapse is characterized by a specific synaptic weight and
transmission delay, accounting for the axonal arborization.
The two-dimensional neural network is mapped on a set of C++ processes 
interconnected with a message passing interface.
Each C++ process simulates the activity of a cluster of neurons.
The spikes generated during the neural activity of each neuron are
delivered to the target synapses belonging to the same or to other
processes.
The ``axonal spikes'', that carry the information about
the spiking neuron identity and the original spike emission time, 
constitute the payload of the messages.
Axonal spikes are sent only toward those C++ processes where a target
synapse exists.

The knowledge of the original emission time and of the 
transmission delay introduced by each synapse is necessary for synaptic Spike Timing Dependent Plasticity (STDP) management, supporting 
Long Term Potentiation/Depression (LTP/LTD) of the synapses. 


\subsection{Execution flow: a mixed time and \mbox{event-driven} approach}
\label{sec:flow}

\begin{figure}[!t]
    \centerline{\includegraphics[width=.50\textwidth]{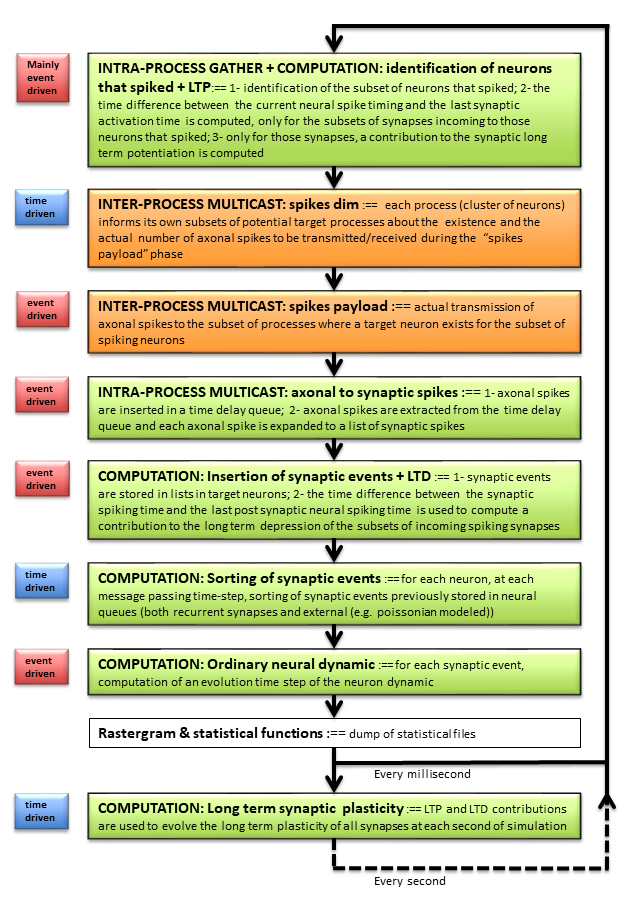}}
    \caption {Execution flow of DPSNN simulator. Labels on the left identify 
event- or time-driven tasks. Orange blocks are used for the inter-processes
communication tasks.}
    \label{fig:flow.png}
\end{figure}

Simulation undergoes two phases:
\setlist[enumerate]{label*=\arabic*.}
\begin{enumerate*}
\item creation and initialization of the network of neurons, of
  the axonal arborization and of the synapses;
\item simulation of the neural and synaptic dynamics.\\

  A combined \mbox{event-driven} and
  \mbox{time-driven} approach has been adopted, inspired
  by~\cite{morrison:2005}: the dynamic of neurons and synapses (STDP) is
  simulated when the event arises (event driven integration), 
  while the message passing conveying the axonal spikes among processes 
  is performed at regular time driven steps (in the present study set to 1 $ms$).
  Simulation (see Fig.~\ref{fig:flow.png})
  can be further decomposed into the following steps:

  \setlist[enumerate]{label*=\arabic*)}
  \begin{enumerate*}

  \item \mbox{spike-producing} neurons during the previous
    \mbox{time-driven} simulation step are identified and the
    consequent contribution to STDP is calculated;
  \item spikes are sent through axonal arborizations to the cluster of
    neurons where target synapses exist;
  \item inside each process, incoming axonal spikes are queued into lists,
    for later usage during the \mbox{time-step}
    corresponding to the synaptic delays;
  \item synapses inject currents into target neurons and the
    consequent contribution to STDP is calculated;
  \item neurons sort input currents coming from recurrent and external
    synapses;
  \item neurons integrate their dynamic equation for each input
    current in the queue, using an \mbox{event-driven} solver.
  \end{enumerate*}

\end{enumerate*}

At a slower timescale, which in the current implementation is every
second, STDP contributions are integrated in a Long Term
Plasticity and applied to each single synapse.

\subsection{Distributed generation of synaptic connections}
\label{sec:conn}

The described simulation engine exploits its parallelism also during
the creation and initialization phase of the network, as detailed in
a following section.
In a given process, a set of local neurons $i = 1,..,N$ projects their set of
synapses $j=1,..,M$, toward their target neurons $K^{i,j}$, each synapse 
characterized by individual delays $D^{i,j}$ and plastic weights $W^{i,j}$.
Synaptic efficacies are randomly chosen from Gaussian distributions, 
while synaptic delays can be generated according to exponential or 
uniform distribution.
The moments of the distributions depend on the source and target
populations that each specific synapse interconnects.
In addition to recurrent synapses, the system simulates also a
number of external synapses: they represent afferent
(thalamo-) cortical currents coming from outside the simulated network.



\subsection{Representation of spiking messages}
\label{sec:aer}

Spike messages are defined using an Address Event Representation
(AER,~\cite{lazzaro2:1993}) including the spiking neuron identifier
and the exact spiking time.
During simulation, spikes travel from the source to the target neuron.
Spikes, whose target neurons belong to the same process, are packed in
the axonal spike message.

The arborization of this message is deferred to the target process.
Deferring as much as possible the arborization of the ``axon'' reduces
the communication load and unnecessary wait barrier.

To this purpose, preparatory actions are performed during network
initialization (performed once at the beginning of the
simulation), to reduce the number of active communication channels
during the iterative simulation phase.

\subsection{Initial construction of the connectivity infrastructure}
\label{sec:init}

During initialization, each process contributes to create
awareness about the subset of processes that should be listened
to during next simulation iterations. This knowledge is based on
information extracted from the locally constructed matrix of
outcoming and incoming synapses.
At the end of this construction phase, each ``target'' process should
know about the subset of ``source'' processes that need to communicate
with it, and should have created its database of locally incoming
axons and synapses.

A simple implementation of the construction phase can be carried out
using two steps.
During the first step, each source process informs other processes
about the existence of incoming axons and the number of incoming
synapses.
A single word, the synapse counter, is communicated among pairs of
processes.
Under MPI, this can be achieved by the \texttt{MPI\_Alltoall()} routine.
This is performed once, and with a single word payload.

The second construction step transfers the identities of synapses to
be created onto each target process.
Under MPI, the payload --- a list of synapses specific for each pair
in the subset of processes to be connected --- can be transferred
using a call to the \texttt{MPI\_Alltoallv()} library function.
The number of messages depends on the lateral connectivity range and
on the distribution of cortical modules among processes, while the
cumulative load is always proportional to the total number of synapses
in the system.

The knowledge about the existence of a connection between a pair of
processes can be reused to reduce the cost of spiking transmission
during the iterations of simulation.

\subsection{Delivery of spiking messages during the simulation phase}
\label{sec:sim}

At each iteration, spikes are exchanged between pairs of processes
connected by the synaptic matrix.
The delivery of spiking messages can be split in two steps, with
communications directed toward subsets of decreasing size.

During the first step, single word messages (spike counters) are sent
to the subset of potentially connected target processes.
On each pair of \mbox{source-target} process subset, the individual
spike counter informs about the actual payload --- \ie axonal spikes
--- to be delivered, or about the absence of spikes to
be transmitted.
The knowledge of the subset was created during the first step of
initialization (see Section~\ref{sec:init}).

The second step uses the spiking counter to establish a
communication channel only between pairs of processes that actually
need to transfer an axonal payload during the current iteration.
On MPI, both steps can be implemented using calls to the
\texttt{MPI\_Alltoallv()} library function.

For \mbox{two-dimensional} grids of neural columns and for
their mapping on processes used in this experiment, this implementation
proved to be quite efficient, as reported in 
Section~\ref{sec:dpsnnresult}, further refined in Section~\ref{sec:dpsnndisc}.
%

\section{Neural Network Configuration}
\label{sec:dpsnnnetwork}

\subsection{Spiking Neuron Model and Synapses}
\label{sec:neuron}

The \mbox{single-compartment}, \mbox{point-like} neurons used
in this paper are based on the Leaky Integrate and
Fire (LIF) neuron model with \mbox{spike-frequency} adaptation (SFA)
due to calcium- and \mbox{sodium-dependent}
\mbox{after-hyperpolarization} (AHP)
currents~\cite{gigante:2007}.
Neuronal dynamics is described by the following equations:

\begin{equation}\label{membrane}
\frac{dV}{dt} = \frac{V-E}{\tau_{m}} - g_{c}\frac{c}{C_{m}} + \sum J_{i}\delta(t-t_{i})
\end{equation}
\begin{equation}\label{fatigue}
\frac{dc}{dt} = - \frac{c}{\tau_{c}}
\end{equation}
where $V(t)$ represents the membrane potential and $c(t)$ the fatigue
variable used to model the SFA as an \mbox{activity-dependent}
inhibitory current.
$\tau_m$ is the membrane characteristic time, $C_m$ the membrane
capacitance, $E$ the resting potential and $\tau_c$ the decay time 
for the fatigue variable $c$. $g_{c}$, paired with the membrane capacitance,
determines the timescales of the coupling of the fatigue \eqref{fatigue} 
and membrane potential \eqref{membrane} equations.
For inhibitory neurons, the SFA term is set to zero.
Synaptic spikes, reaching the neuron at times $t_i$, produce instantaneous 
membrane potential changes of amplitude $J_i$, the weights of activated 
synapses.
When the membrane potential exceeds a threshold $V_\theta$, a spike
occurs.
Thereafter, the membrane potential is reset to $V_r$ for a refractory
period $\tau_{arp}$, whereas the variable $c$ is increased by the
constant amount $\alpha_c$.

During the construction phase of the network, recurrent synapses are
established between pre- and \mbox{post-synaptic} neurons (see
Section~\ref{sec:model}).
Synaptic efficacies and delays are randomly chosen from probabilistic 
distributions (see Section~\ref{sec:conn}).

In addition to the recurrent synapses, the system simulates also a
number of external synapses: they represent afferent
\mbox{(thalamo-)cortical} currents coming from outside the simulated
network, collectively modeled as a Poisson process with a given
average spike frequency.
The recurrent synapses plus the external synapses yield the number of
total synapses afferent to the neuron, referred to as ``total
equivalent'' synapses in the following.

For all the measurements in this work, synaptic plasticity has been
disabled, to simplify the comparison between different 
configurations used in the scaling analysis, ensuring higher stability 
of the states of the networks.

\subsection{Cortical Columns and their connectivity}
\label{sec:model}

Neurons are organized in cortical modules (mimicking columns),
each one composed of 80\% excitatory and 20\% inhibitory neurons.
Modules are assembled in \mbox{two-dimensional} square grids,
representing a cortical area slab, with a grid step $\alpha \sim
100$~$\mu$m (\mbox{inter-columnar} spacing).
The size of these grids has been varied as per Table~\ref{tab:grid}, 
to perform the scalability experiments here reported.

Each cortical module includes 1240 neurons, while
the number of synapses projected by each neuron depends on the
implemented connectivity.

The neural network connectivity is set by the user defining the
probabilistic connection law between neural populations, spatially
located in the \mbox{two-dimensional} grid of cortical modules.
Connectivity can be varied according to the simulation needs, leading
to configurations with different numbers of synapses per neuron.
We adopted the following lateral connectivity rules to evaluate
the impact of different \mbox{inter-module} connectivity laws:
\begin{itemize}
\item Gaussian connectivity --- shorter range and lower number of
  remote synapses: considering preeminent local connectivity with
  respect to lateral, the rule used to calculate remote connectivity
  has been set proportional to $A \cdot exp (\frac{-r^2}{2\sigma^2}$),
  with $A=0.05$ and $\sigma=100~\mu$m being the lateral spread of the
  connection probability. The remote connectivity function is similar
  to that adopted by the~\cite{potjans:2014} model, although with different $A$
  and $\sigma$ parameters. In this case only $\sim$20\% of the
  synapses (specifically $\sim$250) are remotely projected and reach modules placed within a
  short distance, spanning a few steps in the \mbox{two-dimensional}
  grid of cortical modules. The majority of connections ($\sim$80\%)
  is local to the module.
\item Exponential decay connectivity --- longer range and higher
  number of remote synapses: the connectivity rule for remote synapses
  calculation is proportional to $A \cdot exp(\frac{-r}{\lambda}$),
  with $A=0.03$ and $\lambda=290~\mu$m (the exponential decay
  constant, in the range of experimental biological values, see 
  \eg~\cite{schnepel:2015}). This turns out into an increased number 
  of remote connections ($\sim$59\%), \ie $\sim$1400 lateral synapses per neuron.
  It is worth nothing that full biological realism would require to increase
  the total number of lateral connections above $\sim$4~K synapses/neuron.
\end{itemize}

For both studied connectivities, a local connection probability of 80\%
(producing about 990 local synapses) has been adopted.
For classical short-range configuration, the dominance of local synapses
enables mean-field theory prediction of the dynamical regime of the modules,
that perceive the influence of remote modules as small perturbations.
In summary, the average number of projected synapses per neuron is $\sim$2390
for the longer-range exponential connectivity while for the  Gaussian connectivity is $\sim$1240.

\begin{figure}[!b]
    \centerline{\includegraphics[width=.50\textwidth]{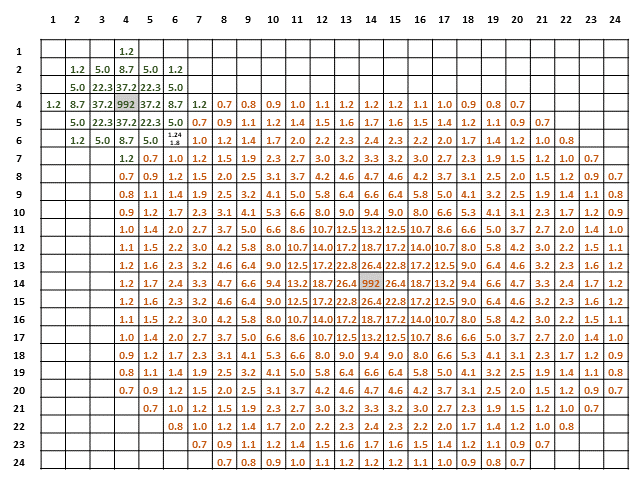}}
    \caption {Example of Gaussian (green) and exponential \mbox{longer-range}
      (orange) connectivity in a grid composed by
      $24 \times 24$ cortical modules: total number of synapses
      (in thousands) projected by excitatory neurons located in the
      column marked in grey. Inhibitory neurons project only local connections.}
    \label{fig:stencil}
\end{figure}

In both systems, a \mbox{cut-off} has been set in the synapses
generation, limiting the projection to the subset of modules with
connection probability greater than 1/1000.
This turns out into stencils of projected connections centered on the source module.
A $7\times 7$ stencil is generated in the first case (Gaussian) and a $21 \times 21$ 
in the second case (exponential decay). They are marked in green and
orange in Fig.~\ref{fig:stencil}.

For each connectivity scheme, measurements were taken on different
problem sizes obtained varying the dimension of the grid of modules
and, once fixed the problem size, distributing it over a span of MPI
processes to evaluate the scaling behaviour.

We selected three grid dimensions: $24 \times 24$, $48 \times 48$ and $96 \times 96$
(see Table~\ref{tab:grid}). For a columnar spacing of few hundreds of microns, 
they can be considered representative of interesting
biological cortical slab dimensions.
The number of processes over which each network size is distributed
varied from a minimum, bounded by memory, and a maximum, bounded by
communication (or HPC platform constraints).

Using the Gaussian \mbox{shorter-range} connectivity, an extensive
campaign of measures has been conducted, spanning over the three
configurations above described.
The impact of \mbox{longer-range} exponential decay interconnects has been
evaluated on the $24 \times 24$ and $48 \times 48$ configurations.

\subsection{Toward biological modeling}
\label{sec:biomodel}
\begin{figure}[!t]
    \centerline{\includegraphics[width=.50\textwidth]{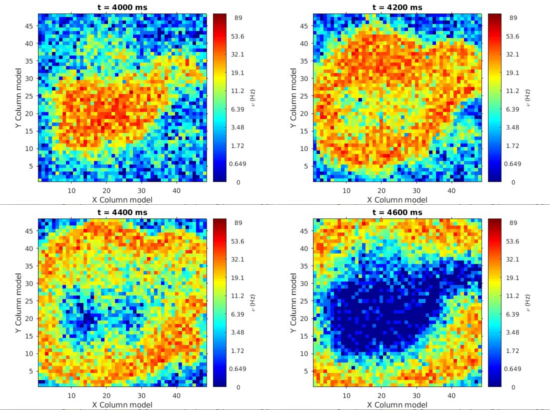}}
    \caption {Four snapshot of a slow wave propagating on a $48 \times 48$ grid of cortical modules spaced at 400~$\mu$m,
with a connection probability exponentially decaying with $\lambda$ = 240~$\mu$m.}
    \label{fig:waves}
\end{figure}

\begin{figure}[!t]
    \centerline{\includegraphics[width=.30\textwidth]{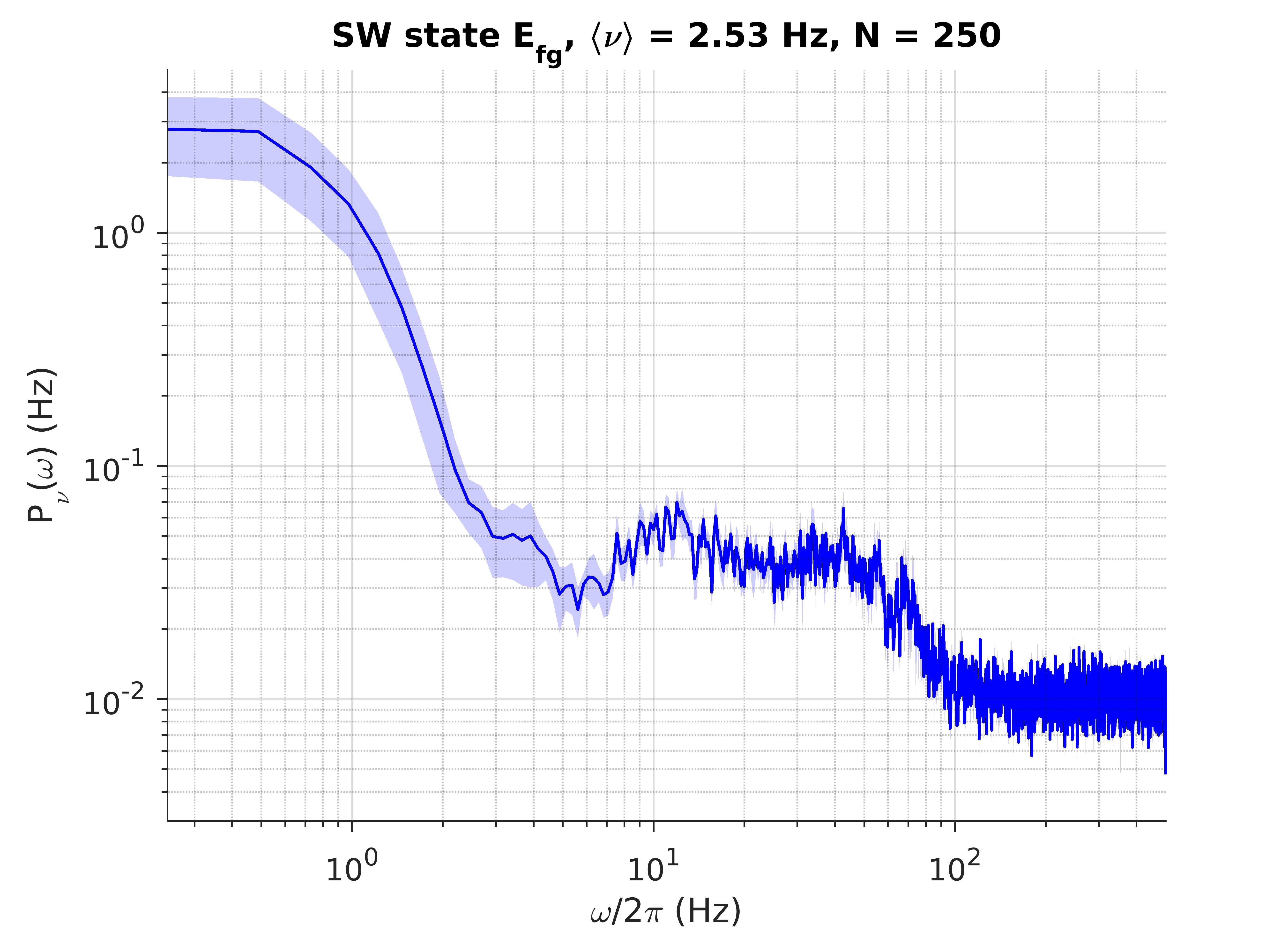}}
    \caption {Power spectral density of a population of excitatory neurons
showing a high quantity of energy in delta band ($<$ 4~Hz).}
    \label{fig:psd}
\end{figure}

The network size and execution speed reached in the reported scaling
experiments makes this engine a valuable candidate tool for the acceleration
of large-scale simulations.
Here, we report a preliminary example of usage
in a specific case of our interest: the modeling of cortical Slow Wave Activity (SWA).
To this purpose, we use a three-dimensional variant of the two-dimensional model 
\cite{caponeSanchezvivesDelGiudiceMattia:2017}.
The development of the variant and its biological meaning will be presented in a 
forthcoming publication (preliminary info in \cite {posterDPSNN:2017}).
Snapshots of an exemplary propagating wave are reported in Fig.~\ref{fig:waves}.
Simulations express delta rhythms, the main feature of SWA,
as shown in their power  spectral density (Fig.~\ref{fig:psd}).
The model includes 2.9~M neurons projecting 3.2~G synapses
arranged in a grid of $48 \times 48$ cortical modules, spaced at 400~$\mu$m,
with a connection probability exponentially decaying with $\lambda$ = 240~$\mu$m.
However, the focus of this paper is on the parallel and distributed 
computing aspects of the engine development and the cost of longer-range lateral 
connectivity.
Papers targeting biological realism are currently under preparation in
cooperation with the partners of the WaveScalES experiment
in the Human Brain Project.  

\subsection{Normalized simulation Cost per Synaptic Event}
\label{sec:synevent}
Different network sizes and connectivity models have been used in this
scaling analysis.
This results in heterogeneous measures of the elapsed time due to
different numbers of projected synapses and to the different firing
rates of resulting models.
For example, the observed firing rate is $\sim$7.5~Hz for the shorter
range connectivity, and in the range between 32 and 38~Hz for
the longer range one (all other parameters being kept constant).
However, a direct comparison is possible converting the execution time
into a simulation cost per synaptic event.
This normalized cost is computed dividing the elapsed time per
simulated second of activity by the number of synapses and by their
mean firing rate.
In this way, a simple comparison among different simulated configurations
is possible: measures from different simulations can be compared on
the same plot.
Our simulations include two kind of synapses: recurrent --- \ie
projected by simulated neurons --- and synapses bringing an external
stimulus.
Summing the number of events generated by recurrent and external
synapses, in the following we can normalize the cost to the total
number of equivalent synaptic events.

\subsection{Hardware Platform}
\label{sec:hwplatform}

We run the simulations on a partition of 64 IBM nodes (1024 cores) 
of the \galileo server platform, provided at the 
\cineca~\cite{cineca:2017:Online} supercomputing center.
Each \mbox{16-core} computational node contains two Intel Xeon Haswell
\mbox{8-core} \mbox{E5-2630} v3 processors, with a clock of 2.40~GHz.
All nodes are interconnected through an InfiniBand network,
equipped with 4$\times$ QDR switches.
Due to the specific configuration of the server platform, no
\mbox{hyper-threading} is allowed.
Therefore, in all following measures, the number of cores
corresponds exactly to the number of MPI processes launched on a 
computational node at each execution.

\section{Results}
\label{sec:dpsnnresult}

\begin{table*}[!hbt]
\caption[Configurations used for the scaling measures of
  \dpsnn.]{Problem sizes for the comparison of simulator performance
  applied to exponential (\mbox{longer-range}) and Gaussian
  (\mbox{shorter-range}) connectivity.}
\label{tab:grid}
\centering
\begin{tabular}{|c|c|c|c|c|c|c|c|c|}
\hline
\textbf{Grid} & \textbf{Columns} & \textbf{Neurons} & \multicolumn{4}{|c|}{\textbf{Number of Synapses}} & \multicolumn{2}{|c|}{\textbf{MPI Procs}} \\ 
\hline
& & & \multicolumn{2}{|c|}{\textbf{Gaussian Connectivity}} & \multicolumn{2}{|c|}{\textbf{Exponential Connectivity}} &  \textbf{Min}  & \textbf{Max} \\
\hline
& & & Recurrent & Total & Recurrent & Total & & \\
\hline
$24 \times 24$   & 576   & 0.7~M   & 0.9~G  & 1.2~G  & 1.5~G  & 1.8~G  & 1  & 64 \\
\hline                   
$48 \times 48$   & 2304  & 2.9~M   & 3.5~G  & 5.0~G  & 5.9~G  & 7.4~G  & 4  & 256  \\
\hline                   
$96 \times 96$   & 9216  & 11.4~M  & 14.2~G & 20.4~G & 23.4~G & 29.6~G & 64 & 1024  \\
\hline
\end{tabular}
\end{table*}

\subsection{Scaling for shorter range Gaussian connectivity}
\label{sec:scaling}

\begin{figure}[!b]
    \centerline{\includegraphics[width=.40\textwidth]{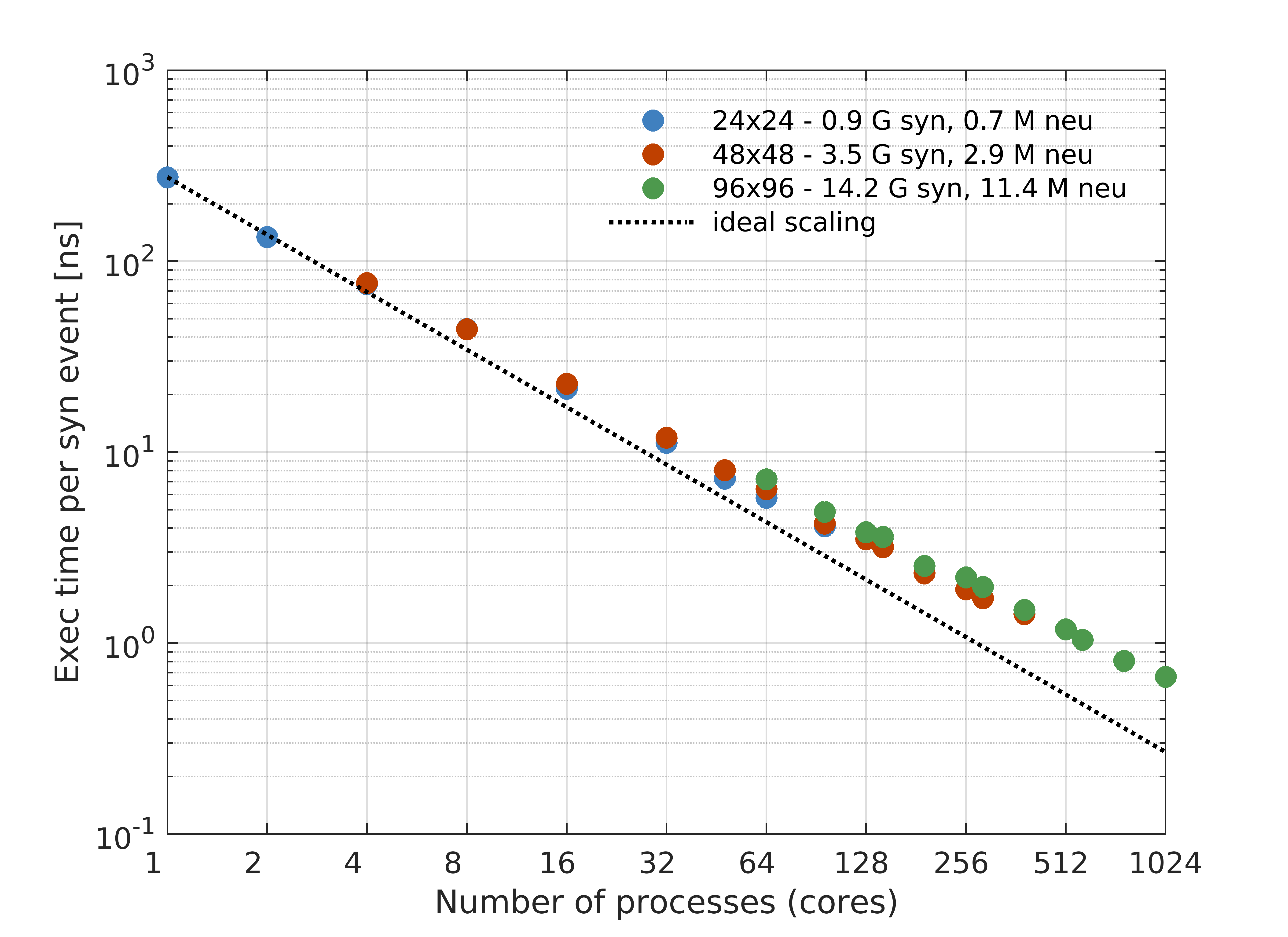}}
    \caption {Strong scaling for Gaussian connectivity model: the
      measures are expressed in elapsed time per equivalent synaptic
      event.}
    \label{fig:strong}
\end{figure}

We collected wall clock execution times simulating different problem
sizes (detailed in Table~\ref{tab:grid}), spanning from 1 to
1024 MPI processes (or, equivalently, hardware cores). Fig.~\ref{fig:strong}
is about the strong scaling of the execution
time per synaptic event.
The black dotted line is the ideal scaling: doubling the resources,
the execution time should halve.
For the $24 \times 24$ grid (0.9~G recurrent synapses and 1.2~G total
equivalent synapses)
the time scales from 275~ns per synaptic event,
using a single core, down to 4.09~ns per event using 96 cores.
The corresponding \mbox{speed-up} is 67.3 times, losing $\sim$30\% compared to the
ideal (96 times).
For the $48 \times 48$ grid (3.5~G recurrent, 5~G
equivalent synapses) the \mbox{speed-up} is 54.2 times (ideal 96 times).
For the $96 \times 96$ grid (14.2~G recurrent/ 20.4~G equivalent
synapses) the \mbox{speed-up} is 10.8 times
(in this case 16 times would be the ideal).

\begin{figure}[!b]
    \centerline{\includegraphics[width=.40\textwidth]{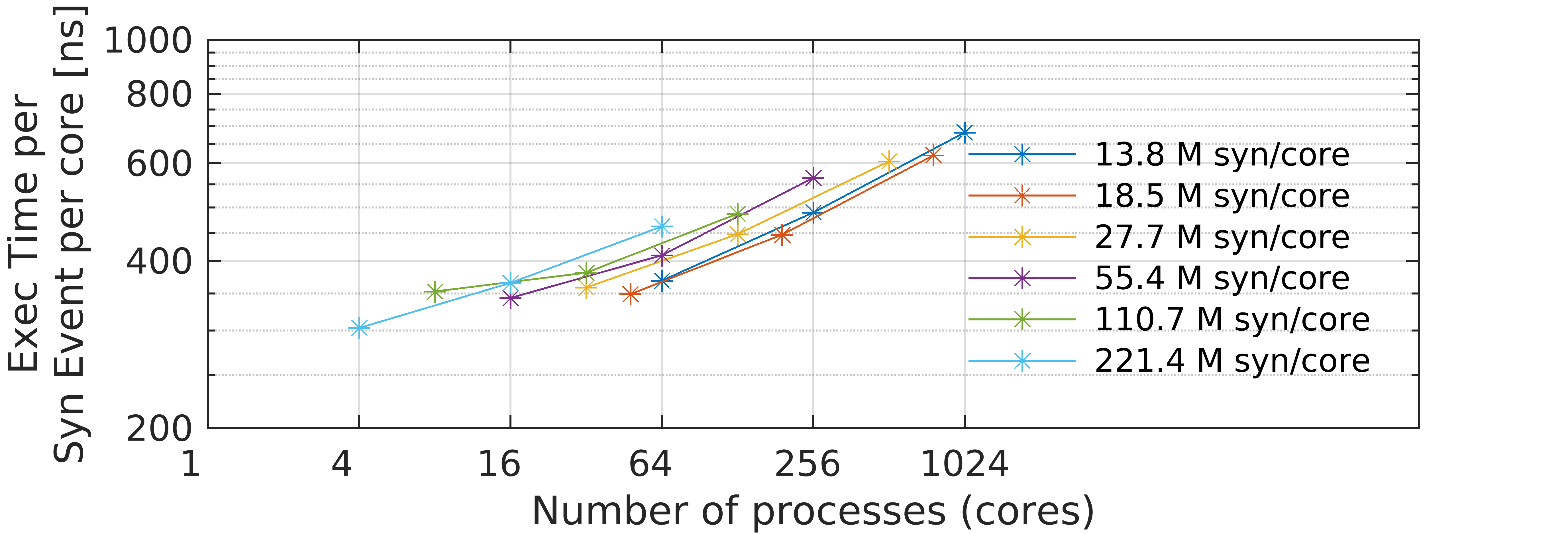}}
    \caption {Weak scaling for Gaussian connectivity model.}
    \label{fig:weak}
\end{figure}

Figure~\ref{fig:weak} reports six curves of weak scaling: constant workload per core,
while increasing the number of resources and the problem size by up to 16 times.
The weak scaling efficiency ranges from 72\% (for a workload of 110.7 M synapses per core)
down to 54\% (when only 13.8 M synapses per core are allocated). 
Ideal weak scaling (100\% efficiency) would produce horizontal lines.
Three points per workload are reported: indeed, data have been extracted from
the run times of the three configurations
$24 \times 24$, $48 \times 48$, $96 \times 96$
used for strong scaling analysis.

In our experience main factors affecting the scaling are collective communications and
timing jitter of individual processes due to both operating system interruptions and
fluctuations in local workload \cite{EURETILE:JSA:2016:short}.

\subsection{Impact of longer range exponential decay connectivity}
\label{sec:impact}

\begin{figure}[!b]
    \centerline{\includegraphics[width=.40\textwidth]{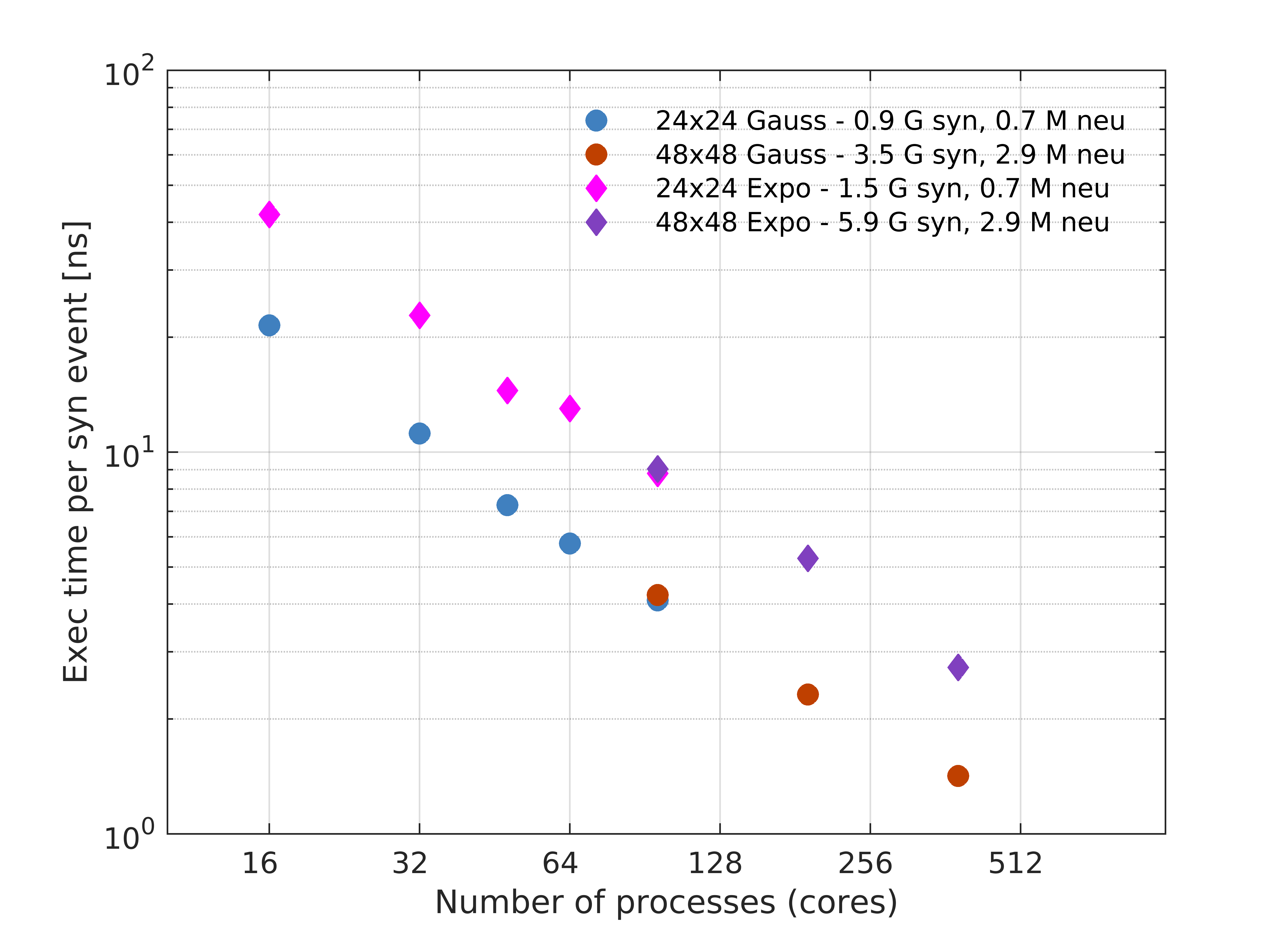}}
    \caption {Impact of lateral connectivity: the graph
      compares the execution time per synaptic event for the
      configurations with Gaussian connectivity (shorter range, lower
      number of synapses - circles) and the one with exponential
      connectivity (longer range, higher number of synapses - diamonds).}
    \label{fig:strongComp}
\end{figure}

Fig.~\ref{fig:strongComp} compares the impact of shorter and longer
lateral connectivity on the strong scaling behaviour.
Circles represent measurements for the Gaussian decay while diamonds
involve the longer range exponential one.

\begin{figure}[!t]
\centering
    \centerline{\includegraphics[width=.40\textwidth]{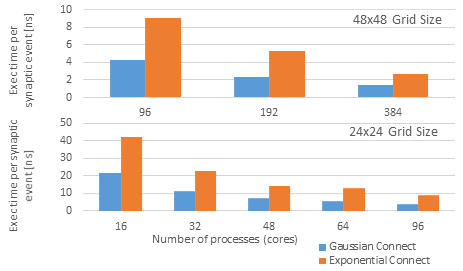}}
    \caption {Time per simulated synaptic event increased between 1.9
      and 2.3 times on changing the decay of connection probability from
      the shorter range Gaussian scheme to the longer range
      exponential one.}
    \label{fig:timeComp}
\end{figure}

The introduction of longer range connectivity increases the simulation
cost per synaptic event, with a \mbox{slow-down} between $1.9$ and $2.3$ times, 
(see Fig.~\ref{fig:timeComp}).
The actual elapsed simulation time increased up to 16.6 times for the
exponential \mbox{longer-range} connectivity due to the combined
effect produced by: (i) the number of synapses projected by each
neuron is higher (by a factor of 1.65), (ii) the firing rates expressed
by the model is between 4.3 and 5.0 times higher and (iii) the higher
cost of longer range communication and demultiplexing neural spiking
messages.
Point (iii) should be well estimated by the \mbox{slow-down} of the
normalized simulation cost per synaptic event.
The execution of longer range connectivity on 96 cores reached about
83\% for the $48\times48$ (5.9~G recurrent synapses) and 79\% of the
ideal scaling for the $24\times24$ case (1.5~G recurrent synapses).


\subsection{Memory cost per synapse}
\label{sec:memory}

\begin{figure}[!t]
    \centerline{\includegraphics[width=.40\textwidth]{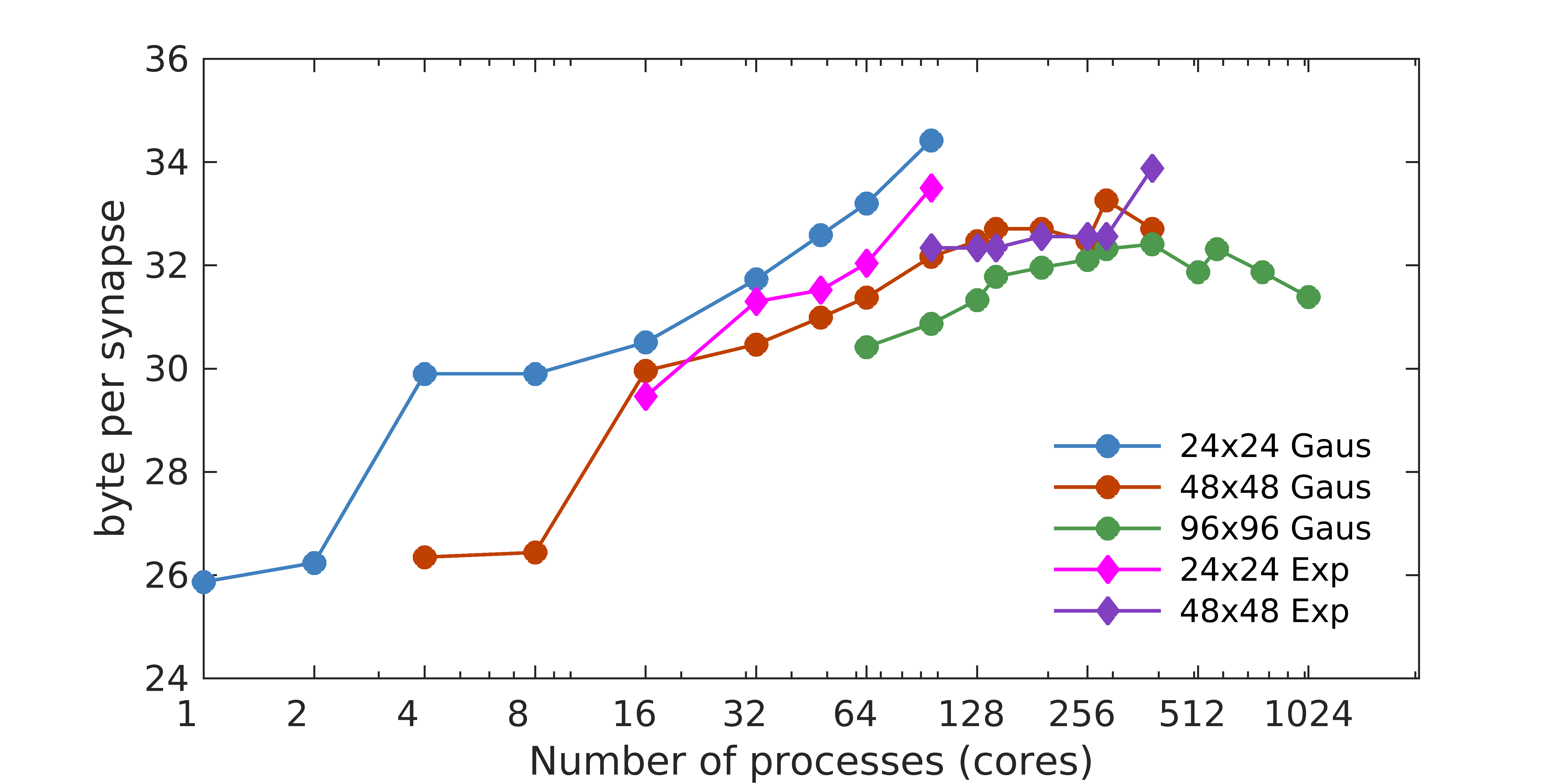}}
    \caption {Memory occupation in byte per
      synapse for different configurations in the two connectivity
      systems}
    \label{fig:memPerSyn}
\end{figure}

We measured the total amount of memory allocated and
divided it by the number of represented synapses.
With no plasticity, each synapse should cost 12~Byte.
Peak memory usage is observed at the end of initialization,
when each synapse is represented at both the source and target
process.
Afterwards, memory is released on the source process.
The forecast of minimum peak cost is therefore 24~Byte/synapse for
static synapses.
Fig.~\ref{fig:memPerSyn} shows the maximum memory footprint for
different networks sizes and projection ranges, distributed over
different numbers of MPI processes.
The values are in the range between 26 and 34~Byte per synapse.
We observed that the growing cost for higher number of MPI processes
is mainly due to the memory allocated by the MPI libraries.

\section{Discussion}
\label{sec:dpsnndisc}
Recent experimental results suggest the need of supporting long range
lateral connectivity in neural simulation of cortical areas --- \eg
modeled by simple exponential decay of the connection probability ---
with layer to layer specific decay constants, in the order of several
hundreds of microns.
A distributed spiking neural network simulation engine (\dpsnn) has been applied
to \mbox{two-dimensional} grids of neural columns spaced at 100~$\mu$m
connected using two schemes.

The \mbox{longer-range} connectivity model corresponds to an
exponential connectivity decay ($\lambda=290~\mu$m) and to the
projection of approximately $\sim$2390 synapses per neuron.
The performance of the engine is compared to that obtained with a shorter
range Gaussian decay of connectivity, with a decay constant of the
order of the columnar spacing and a lower number of synapses per
neuron ($\sim$1240).
The impact of  \mbox{longer-range} \mbox{intra-areal} exponential
connectivity is indeed observable: it increases the simulation cost
per synaptic event between 1.9 and 2.3 times compared to
traditional \mbox{shorter-range}.
The trends of the scaling are quite similar for the two studied connectivities.   
Notwithstanding the slow-down due to longer range connectivity,
the engine demonstrates the ability to simulated large grids of
neural columns (up to $96 \times 96$), containing a total of up to 11.4~M LIF neurons with
\mbox{spike-frequency} adaptation, and representing up to 30~G equivalent
synapses on a 1024 core execution platform, with a memory occupation always below
35~Byte/synapse. 
This is enough to simulate, on clusters of moderate size,
cortical slabs with \mbox{long-range} \mbox{intra-areal} lateral interconnect,
enabling the modeling of cortical slow waves,
a first objective of our team. 

A second objective of \dpsnn is to support the
hardware-software co-design of architectures dedicated to
neural simulation.
In perspective, we note that more detailed biological simulations of
cortical areas could require further extensions of lateral connectivity
models and the support of more complex connection motifs at
different spatial scales.
A further element in future whole brain simulations
will be the co-design with white matter \mbox{inter-areal} connectome,
which brings sparser connections at system scale. 
A balance between approaches focalizing on sparse connectivity 
like~\cite{kunkel:2014:short} and those considering spatial localization
(like the one adopted by \dpsnn) will have to be carefully addressed
for efficient multiscale simulations
of the whole brain.
The results here presented, combined with previous experiences related to
jitter of execution times of individual processes and the impact of
collective communications when profiling \dpsnn execution on distributed platforms,
jointly suggest the utility of designing improved
hierarchical communication infrastructures for spiking messages,
mechanisms of synchronization of computing nodes and dedicated hardware accelerators.
The improvements should consider requirements imposed by
biological connectivity, at least for
those engines that adopt mapping strategies of neurons and incoming
synapses based on spatial contiguity.
In such a context, \dpsnn
can be used to measure the impact of improved designs of execution platforms.

\section*{Acknowledgment}

This work has received funding from the European Union
Horizon 2020 Research and Innovation Programme under
\hbpsgafirst and under \exanestgrant.
Simulations have been performed on the Galileo platform, 
provided by CINECA in the frameworks of HBP SGA1 and of the 
INFN-CINECA “Computational theoretical physics 
initiative” collaboration. We acknowledge G. Erbacci (CINECA) and L. Cosmai 
(INFN) for the platform setup support.


\bibliographystyle{ieeetr}

\end{document}